\documentclass[]{spie}  

\usepackage{amsmath,amsfonts,amssymb}
\usepackage{graphicx}
\usepackage[colorlinks=true, allcolors=blue]{hyperref}

\title{Sensor fusion on MagAO-X: real time vibration control using accelerometers}

\author[a,b]{Parker T. Johnson}
\author[a]{Jared R. Males}
\author[c]{Povilas Palunas}
\author[e,a,d]{Olivier Guyon}
\author[f,a]{Sebastiaan Haffert}
\author[g]{Joseph Long}
\author[e,h]{Vincent Deo}
\author[e]{Julien Lozi}
\author[a]{Laird M. Close}
\author[a]{Maggie Kautz}
\author[a,b,d]{Jay Kueny}
\author[a,b]{Jialin Li}
\author[a,d]{Joshua Liberman}
\author[a,e]{Miles Lucas}
\author[f]{Matthijs Mars}
\author[a,b,d]{Eden McEwen}
\author[d]{Tiffany Nguyen}
\author[f]{Elena Tonucci}
\author[a,d]{Katie Twitchell}
\affil[a]{Steward Observatory, University of Arizona, 933 N Cherry Ave, Tucson, AZ 85721, USA}
\affil[b]{National Science Foundation Graduate Research Fellow}
\affil[c]{Las Campanas Observatory, La Serena, Chile}
\affil[d]{James C. Wyant College of Optical Sciences, University of Arizona, 1630 E. University Blvd., Tucson, AZ 85721, USA}
\affil[e]{Subaru Telescope, National Astronomical Observatory of Japan, National Institutes of Natural Sciences, 650 North A’ohōkū Place, Hilo, HI 96720, USA}
\affil[f]{Leiden Observatory, Leiden University, PO Box 9513, 2300 RA Leiden, The Netherlands}
\affil[g]{Center for Computational Astrophysics, Flatiron Institute, 162 5th Ave, New York, NY}
\affil[h]{Optical Sharpeners SAS, France}

\authorinfo{Further author information: (Send correspondence to P.T.J.) \\P.T.J.: E-mail: parkertjohnson1@arizona.edu
}

\begin{document} 
\maketitle
\begin{abstract}

Mechanical vibrations are a significant source of residual wavefront error (WFE) in adaptive optics (AO) systems, limiting the performance of high-contrast imaging instruments. We present the design and on-sky deployment of a low-cost, modular accelerometer telemetry system for the MagAO-X extreme AO instrument on the 6.5 m Magellan Clay Telescope, consisting of piezoelectric accelerometers and a Raspberry Pi-based acquisition system that streams synchronized data to the real-time control computer with microsecond-level timing stability. The system is used to identify dominant telescope vibration sources and quantify their coupling to AO telemetry, revealing that several narrow-band modes originate from subsystems including the primary mirror glycol pump, secondary mirror actuation system, and telescope autofocus system. Coherence analysis between the synchronized accelerometer and wavefront sensor telemetry demonstrates that approximately one-third of the residual tip and tilt WFE is correlated with structural vibrations, indicating that accelerometer telemetry provides a promising foundation for future predictive control implementations. These results demonstrate that low-cost accelerometer telemetry provides a practical approach for vibration identification and a foundation for predictive control in current and future AO systems.

\end{abstract}


\section{INTRODUCTION}
\label{sec:intro}  

In ground-based adaptive optics (AO) systems, residual wavefront error (WFE) is driven by both atmospheric turbulence and mechanical vibrations from telescope and instrument structures~\cite{Lozi18, Jaufmann25}. These structural disturbances originate from telescope drives, cooling pumps, fans, mirror support systems, and wind forces\cite{Jaufmann24,Jaufmann25,Lozi18}. All of these vibrations propagate through the telescope structure and directly impact the optical path. Vibrations produce low-order aberrations such as tip, tilt, and focus errors that degrade image quality, reduce coronagraphic suppression, and limit the achievable contrast for AO systems\cite{Lozi18,Bohm2014}. For extreme AO instruments designed for high-contrast imaging (HCI), nanometer scale mechanical motion can translate into milliarcsecond image motion and cause significant stellar leakage\cite{Martinez12}.

Modern AO systems use stellar photons measured by the WFS to correct both atmospheric turbulence and structural vibrations. This creates a fundamental limitation because the WFS bandwidth and signal-to-noise ratio depend strongly on guide star brightness causing HCI AO systems to operate the control loop at slower speeds for fainter stars\cite{Guyon05}, hindering WFE corrections. Accelerometers provide an independent measurement of structural vibrations without consuming stellar photons. This allows the WFS to devote its limited bandwidth to correcting atmospheric turbulence.

Vibrational modal characterization and simulations have been performed on many major ground-based telescopes including the Very Large Telescope~\cite{Jaufmann24,Jaufmann25}, the Subaru Telescope~\cite{Lozi18}, and the Large Binocular Telescope~\cite{Bohm2014}. These studies show that structural vibrations are a consistent limitation for AO systems and will become increasingly important for future Extremely Large Telescope (ELT) class observatories. Larger telescope structures introduce increased mechanical complexity, larger moving masses, and stronger wind forces, all of which are expected to generate additional vibration modes. Wind shake in particular is expected to become a dominant disturbance source due to the large surface area of ELT structures and secondary support systems. This wind load effect on an ELT class telescope is predicted to be an order of magnitude higher than the atmospheric effect\cite{Kan06,Sedghi10}.

In this work, we investigate the use of accelerometers as an additional telemetry source for real-time vibration control within the MagAO-X instrument. By directly measuring structural vibrations on the Magellan Clay telescope, accelerometers provide an independent measurement of mechanical disturbances before they fully propagate into the optical system\cite{Agapito11}. This enables synchronized vibration measurements that can ultimately support predictive control that is independent of guide star brightness and can operate at much higher sampling rates than the WFS alone.

We developed a low-cost, modular, vibration sensing system capable of synchronized high-speed telemetry streaming directly into the MagAO-X real-time control computer (RTC). Accelerometers were mounted on the back of the secondary mirror to identify vibrations correlated with WFS residuals. We then investigated the relationship between structural vibrations and AO telemetry through coherence analysis.

This paper is organized as follows: Section~\ref{sec:system} describes the accelerometer hardware, synchronization architecture, and telemetry pipeline. Section~\ref{sec:results} presents synchronization performance, identified vibration sources, and coherence measurements between accelerometer and WFS telemetry. Section~\ref{sec:discussion} discusses the implications of these results for current and future AO systems, including ELT-class instruments such as GMagAO-X\cite{Males24} on the GMT.

\section{System Overview}
\label{sec:system}

\subsection{Accelerometer Hardware}
\label{sec:hardware}

The vibration sensing system is built to deliver a low noise, high sampling rate, and low-cost solution to measure structural disturbances. At the core of the system is a Raspberry Pi 5, housed in an aluminum alloy passive cooling case. Passive cooling was selected to eliminate any fan induced vibrations that could contaminate the accelerometer measurements and shake the telescope. The Raspberry Pi 5 provides a 4 core processor with 8GB of RAM. We will show that this level of compute is capable of handling real time data acquisition and communication with an AO system. 

Vibration measurements are obtained using a PCB Model 393B05 piezoelectric accelerometer. This sensor is a high sensitivity (10 $\mathrm{V/g}$) with a broadband resolution of $4 \mathrm{~\mu g}$ RMS and a measurement range of $0.5 \mathrm{~g ~pk}$. The accelerometer has built in electronics that convert a high impedance charge signal into a low impedance voltage signal allowing the signal to be transmitted over long coaxial cables with minimal noise. It also converts the raw charge signal from the piezoelectric material into a usable AC voltage signal. We use a PCB Model 482C05 four channel signal conditioner to supply the piezoelectric accelerometers with the necessary constant current excitation and excitation voltage. The multi-channel feature enables simultaneous conditioning of multiple accelerometers while keeping phase consistent measurements across all channels.

Digitization of the analog signals is done using a Microchip MCP3208, which is a 12-bit analog-to-digital converter (ADC) with a serial-peripheral interface (SPI). The MCP3208 provides eight single-ended input channels, allowing scalability for future expansions of the sensing network. The 12-bit resolution offers sufficient quantization precision for capturing sub-micron structural vibrations when combined with appropriate signal scaling. 

\begin{figure}[!ht]
\centering
\includegraphics[width=.5\columnwidth]{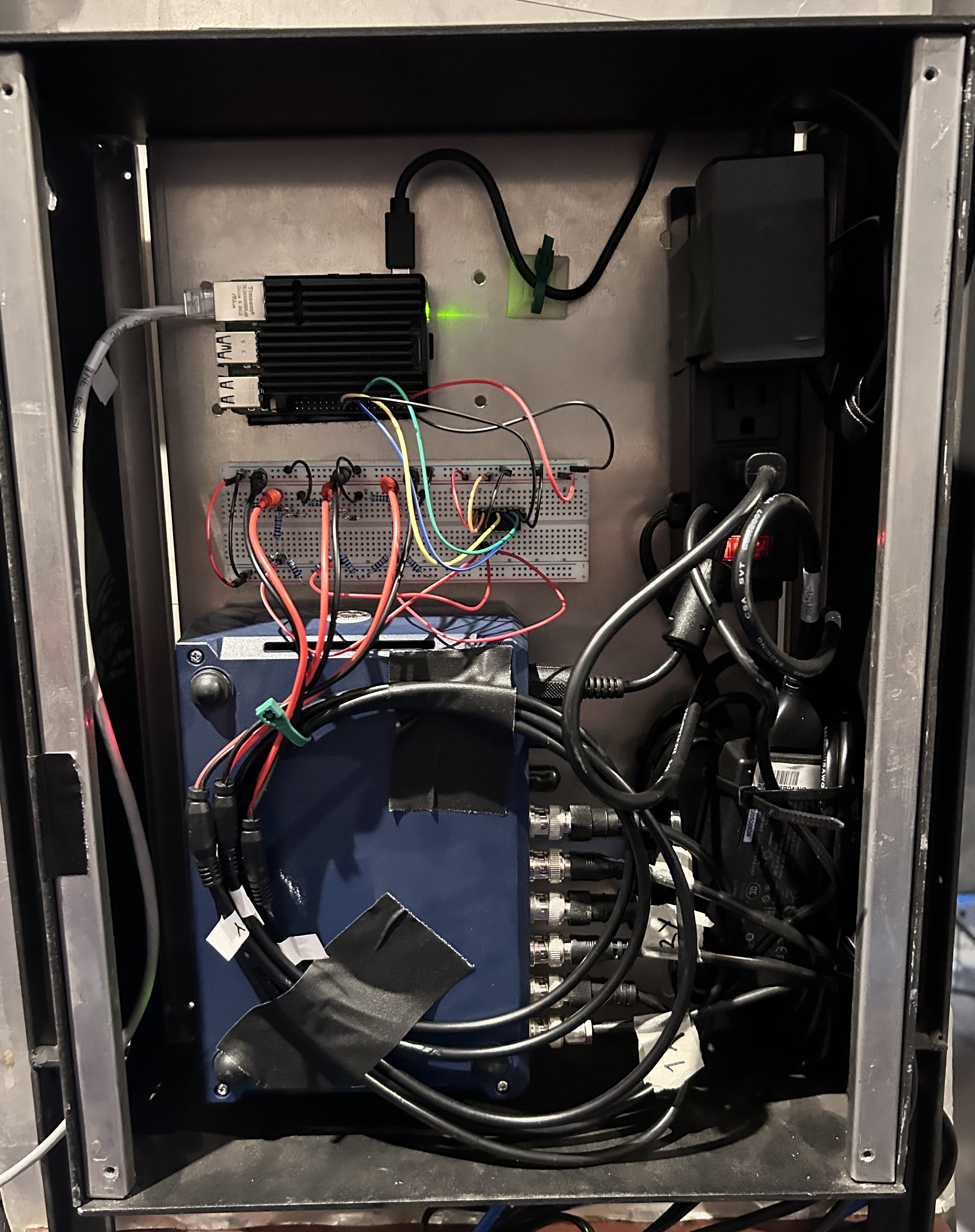}
\vspace{2mm}
\caption{
Electronics enclosure mounted on the back of the telescope top-end structure, housing the accelerometer data acquisition system including the Raspberry Pi 5, signal conditioning, and digitization electronics.
}
\label{fig:electronic_box}
\end{figure}

To protect against transient voltage spikes or potential high voltage conditions, like earthquakes, we designed and implemented a custom voltage divider and clamping circuit  prior to the ADC input stage. This circuit limits the input voltage to within the safe operating range of the MCP3208. The combination of these components allow for the low noise, high sampling rate, and low cost modular vibration sensing system capable of continuous structural monitoring. Figure~\ref{fig:electronic_box} shows the actual layout of the electronics box including the Raspberry Pi 5, signal conditioner, and breadboard with the circuit and ADC. 

\subsection{Accelerometer Network Layout}

\begin{figure}[!ht]
\centering
\includegraphics[width=.95\columnwidth]{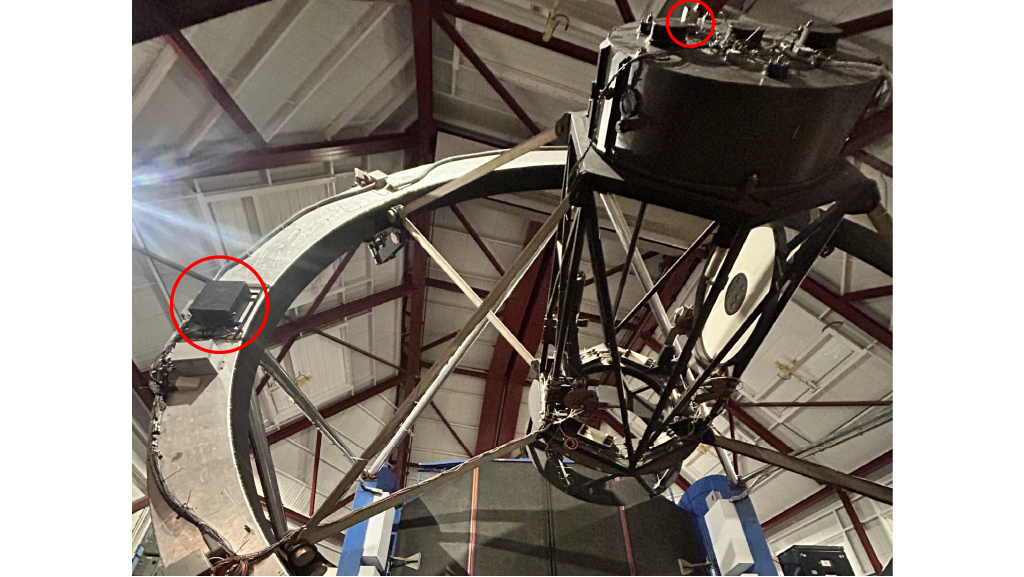}
\caption{
Mounting locations of the vibration sensing system on the telescope structure. The electronics enclosure (left, circled) is installed on the back of the top-end structure, while the accelerometers (right, circled) are mounted on the rear surface of the secondary mirror.
}
\label{fig:accel_electronics_pos}
\end{figure}

\begin{figure}[!ht]
\centering
\includegraphics[width=.5\columnwidth]{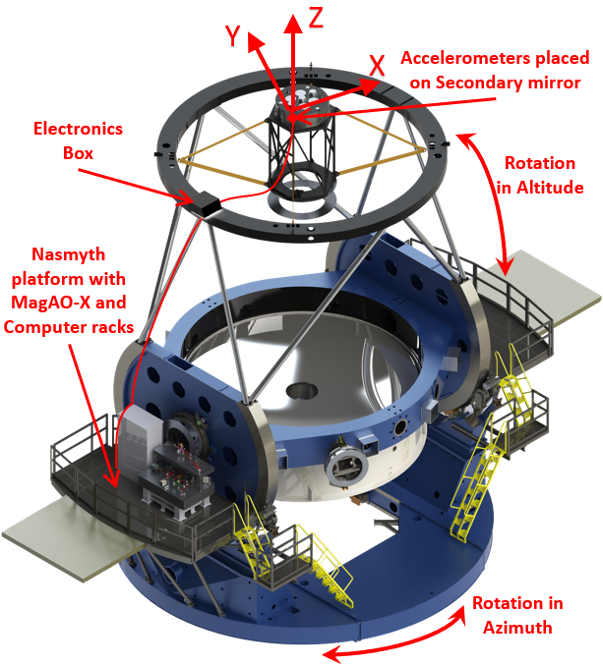}
\vspace{2mm}
\caption{
Rendering of the Clay Telescope with the MagAO-X instrument on the Nasmyth platform. The locations of the accelerometers and electronics enclosure are indicated. A dedicated Ethernet link connects the Raspberry Pi 5 acquisition system directly to the RTC computer on the platform.
}
\label{fig:clay_magaox_rendering}
\end{figure}

The accelerometer network's main purpose is to provide synchronized vibration measurements between the telescope structure and the MagAO-X's AO telemetry streams. Figure~\ref{fig:accel_electronics_pos} and Figure~\ref{fig:clay_magaox_rendering} show the layout of the accelerometer system on the Clay telescope. The electronics enclosure is mounted directly on the back of the top end telescope structure to minimize cable lengths between the accelerometers and the acquisition electronics. The accelerometers are mounted on the back of the secondary mirror where structural disturbances are suspected to be strongly coupled to the optical path and directly impact wavefront stability.

The acquisition system communicates with the MagAO-X RTC, located on the Nasmyth platform, through the observatories Wi-Fi. A direct point-to-point Ethernet connection is used for the real-time data transfer between the Raspberry Pi 5 and the RTC to minimize network latency and prevent congestion from shared observatory network traffic.

Synchronization between the accelerometer acquisition system and the AO telemetry streams is achieved through external timing signals generated by the RTC. The Raspberry Pi receives semaphores from the RTC that initiate an accelerometer readout. This allows the system to maintain temporal alignment between accelerometer measurements, WFS telemetry, and DM commands. Accurate synchronization is a critical component for correlating structural vibrations with residual WFE and enabling future accelerometer-assisted predictive control.

The network architecture is designed to be modular and scalable. Additional accelerometers can be integrated into the system through their own acquisition units while maintaining synchronization with the WFS timings. This allows for future expansions if there are other sources of vibration identified on the telescope or MagAO-X.

\section{Results}
\label{sec:results}

\subsection{Synchronization Performance}

Accurate synchronization is essential for directly comparing accelerometer and WFS telemetry and establishes the foundation for future accelerometer-assisted predictive control. Temporal misalignment reduces the correlation between the telemetry streams, limiting both vibration characterization and future predictive control performance. To quantify the synchronization performance and timing stability of the acquisition pipeline, we measure frame-to-frame timing jitter for three timestamped streams: the local accelerometer readout (ACC-local), where timestamps are recorded immediately after acquisition on the Raspberry Pi; the accelerometer stream on the RTC (ACC), where the same samples are timestamped after transmission to the RTC; and the WFS stream, which is timestamped on the RTC. Comparing ACC-local and ACC isolates the timing uncertainty introduced by communication between the Raspberry Pi and the RTC. Timing jitter is computed as the deviation of frame-to-frame intervals from the nominal 500 $\mu$s (2 kHz) cadence, with only adjacent frames where the counter increment is equal to one are included in the analysis.

We compare two software architectures. The first is a semaphore-based synchronization where a semaphore is sent from RTC at the end of a WFS read, to the Raspberry Pi to trigger an accelerometer read. The second uses an internal integrator-based timing loop to acquire accelerometer samples at 2 kHz. Rather than relying on a fixed-delay loop, the controller accumulates timing error between the measured and desired sampling intervals and adjusts subsequent trigger intervals to maintain the target sampling frequency. Both are evaluated with and without CPU isolation for the shared memory image transmission control protocol communication thread used by the Compute and Control for Adaptive Optics real-time software framework~\cite{Guyon20}. 

\begin{figure}[!ht]
\centering
\includegraphics[width=\columnwidth]{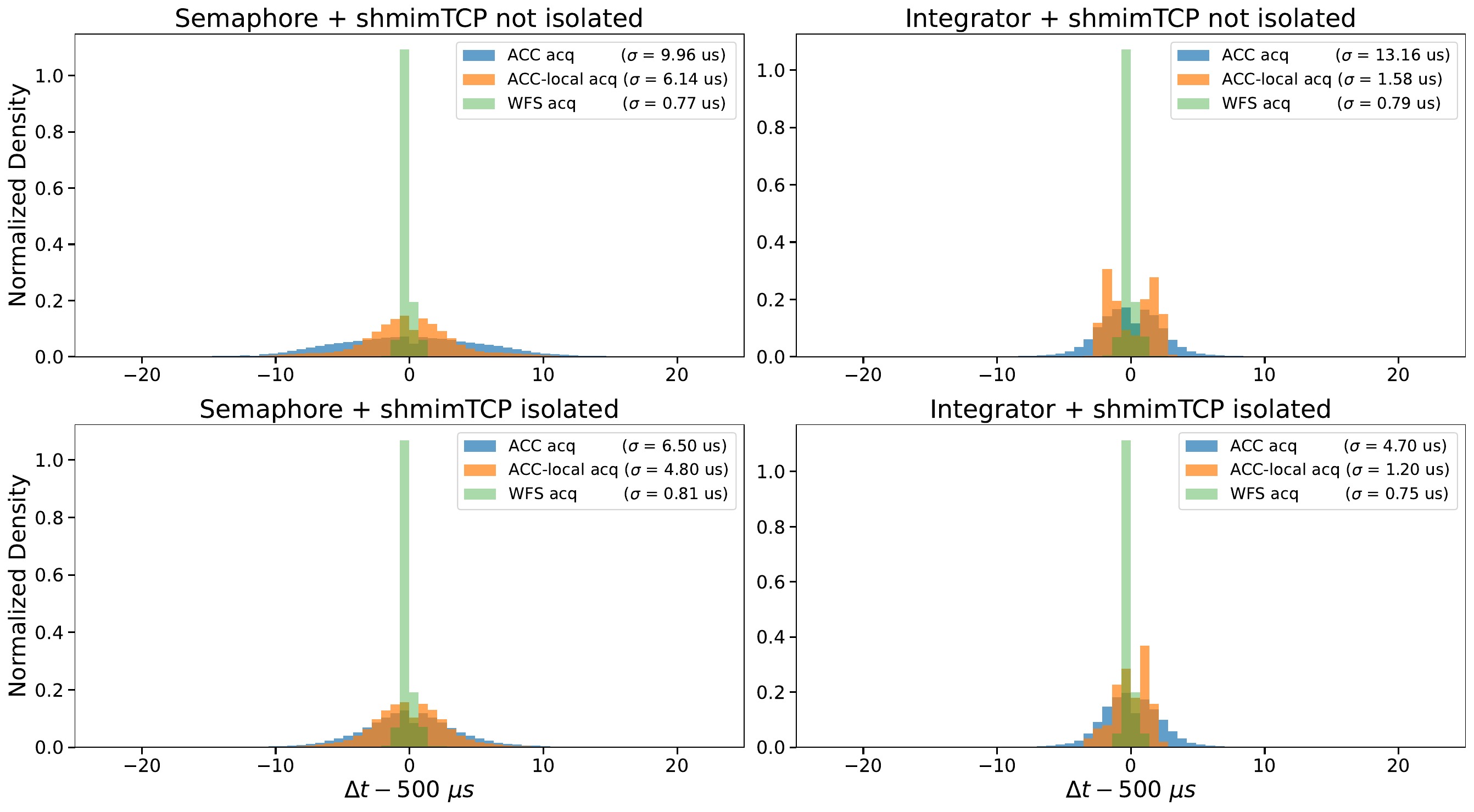}
\caption{
Histogram of frame-to-frame timing jitter for ACC, ACC-local, and WFS streams across semaphore and integrator architectures, with and without CPU isolation. The distributions quantify deviations from the nominal 500 $\mu$s cadence.
}
\label{fig:synchro_histo}
\end{figure}

As expected, the WFS exhibits the lowest timing jitter, consistently below 1 $\mu$s RMS, because its timestamps are generated directly by the RTC. In semaphore mode, ACC-local jitter is dominated by the variability of semaphore delivery and thread wake-up on the Raspberry Pi, where the internally timed integrator avoids this source of latency and therefore produces lower local timing jitter. The ACC measurements have additional timing variability relative to ACC-local because they measure the complete acquisition pipeline, including communication between the Raspberry Pi and the RTC. This additional variability is most significant for the non-isolated integrator configuration, suggesting that software scheduling and communication latency contribute meaningfully to the end-to-end timing jitter. Enabling CPU isolation substantially reduces the ACC jitter for both triggering schemes, demonstrating that operating system scheduling effects is a primary contributor to timing variability. The resulting timing distributions are shown in Figure~\ref{fig:synchro_histo} and summarized in Table~\ref{tab:jitter_summary}.

\begin{table}[ht]
\centering
\caption{Frame-to-frame timing jitter (RMS in $\mu$s) for ACC, ACC-local, and WFS timings in different configurations.}
\label{tab:jitter_summary}
\begin{tabular}{lccc}
\hline
Configuration & ACC & ACC-local & WFS \\
\hline
Semaphore (no isolation)     & 9.96  & 6.14  & 0.77 \\
Semaphore (isolated)        & 6.50  & 4.80  & 0.81 \\
Integrator (no isolation)   & 13.16 & 1.58  & 0.79 \\
Integrator (isolated)       & 4.70  & 1.20  & 0.75 \\
\hline
\end{tabular}
\end{table}

\subsection{Identified and Mitigated Vibrations}

\begin{figure}[!ht]
\centering
\includegraphics[width=\columnwidth]{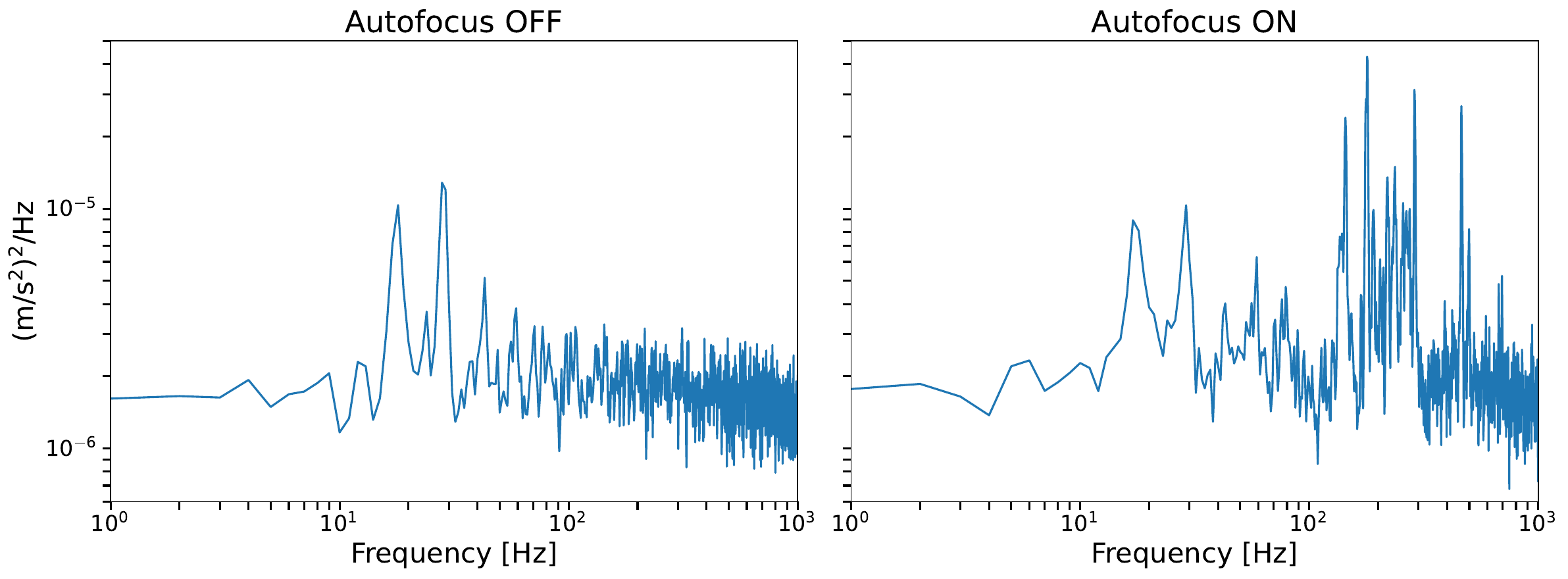}
\caption{
Accelerometer PSDs measured with the telescope autofocus system enabled and disabled. Several vibration peaks appear between 100 and 500 Hz when the autofocus system is active. 
}
\label{fig:autofocus_on_off}
\end{figure}

Figure~\ref{fig:autofocus_on_off} compares the measured accelerometer PSDs with the telescope autofocus system enabled and disabled. Activating the autofocus system introduces several prominent vibration modes between approximately 100 and 500 Hz. This indicates that the autofocus system is a significant source of structural disturbances within the telescope.

\begin{figure}[!ht]
\centering
\includegraphics[width=.9\columnwidth]{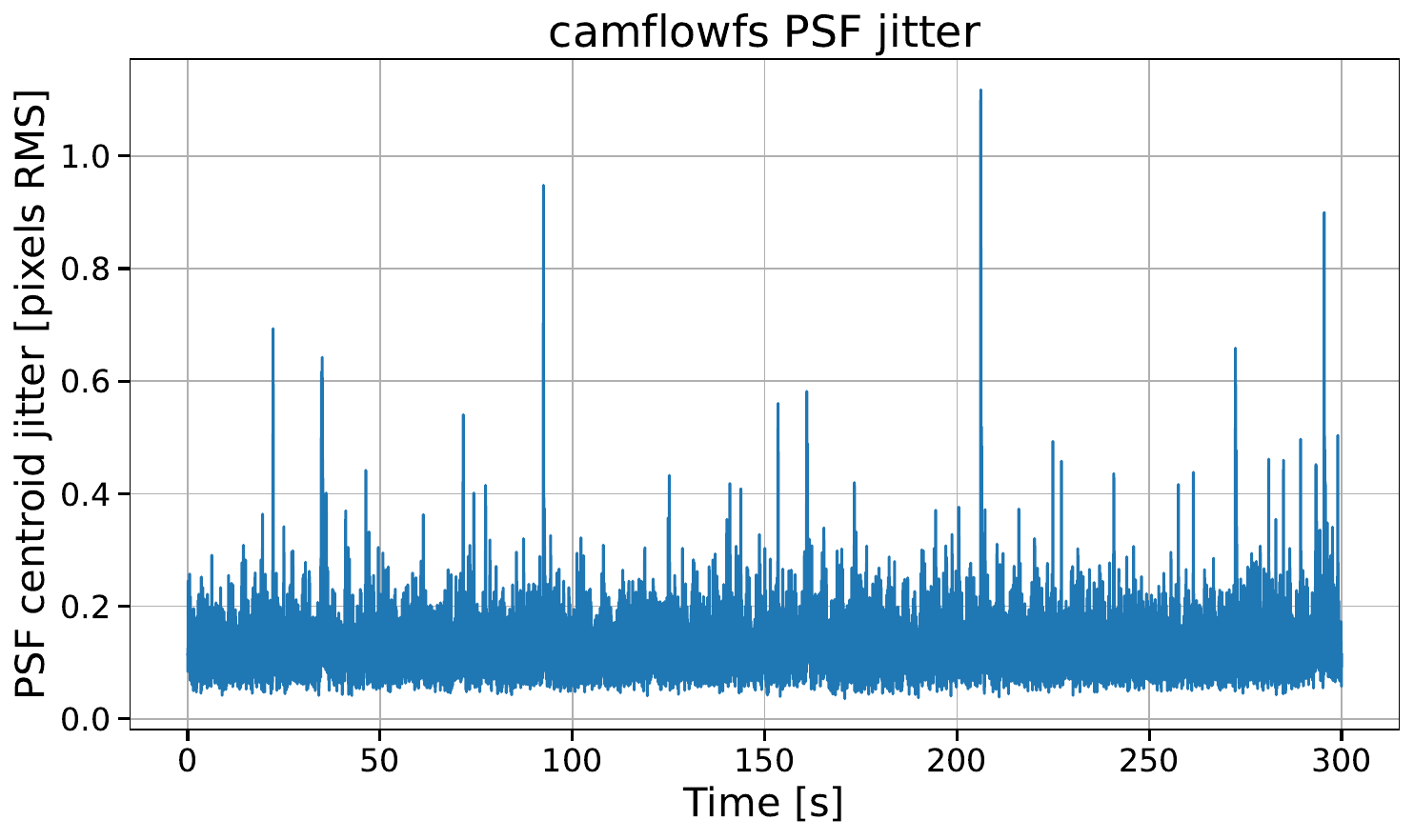}
\caption{
Time series of the camflowfs PSF centroid jitter during telescope autofocus operation. Periods when the autofocus system is active correspond to a significant increase in image motion. 
}
\label{fig:camflowfs_jitter}
\end{figure}

To determine whether these vibrations propagate into the optical system, telemetry from MagAO-X's focal-plane low-order WFS camera\cite{Miller18} (camflowfs) was analyzed. Figure~\ref{fig:camflowfs_jitter} shows the measured PSF centroid jitter during autofocus operation. During autofocus operation, a substantial increase in image motion is observed, with the RMS PSF jitter increasing by nearly an order of magnitude relative to operation with the autofocus system disabled. This demonstrates that vibrations generated by the autofocus system couple directly into the science beam and contribute to residual low-order WFE.

\begin{figure}[!ht]
\centering
\includegraphics[width=\columnwidth]{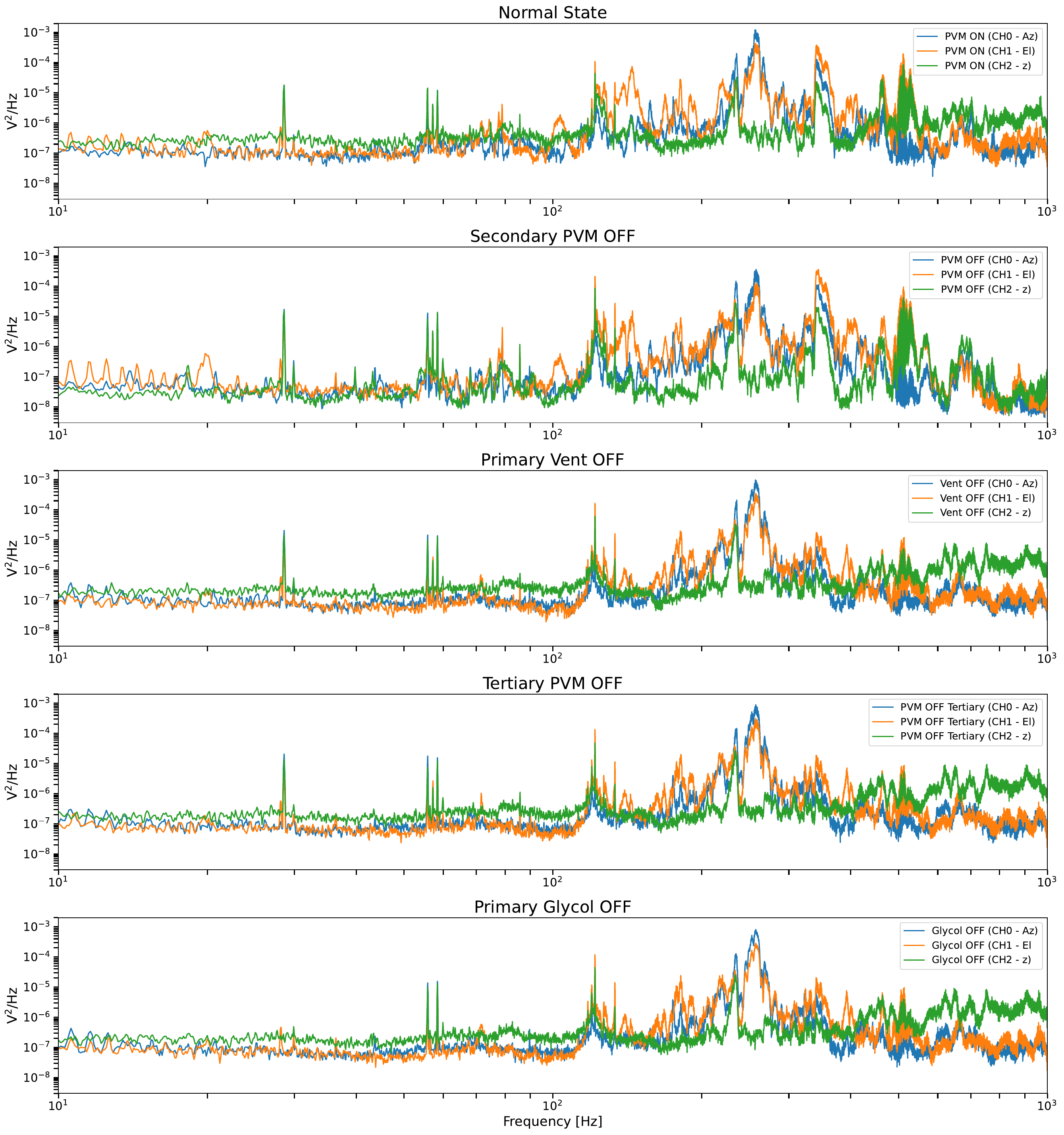}
\caption{
Identifying vibrations through turning telescope functions on and off. The major peaks around 28 Hz and 56 Hz stem from the glycol pump for the primary mirror.
}
\label{fig:tele_vib_test_psds}
\end{figure}

Figure~\ref{fig:tele_vib_test_psds} shows the PSDs of the telescope in a variety of different configurations. The first plot is with all telescope systems running to get a reference PSD. The second plot shows the PVM system off on the secondary mirror. The third plot is with the vent off on the primary mirror. The fourth is with the PVM off on the tertiary mirror. The final plot shows the PSD with the primary glycol pump off. This result identifies the primary mirror glycol pump as the source of the prominent 28 Hz and 56 Hz vibration peaks observed in both the accelerometer and WFS telemetry.

\begin{figure}[!ht]
\centering
\includegraphics[width=\columnwidth]{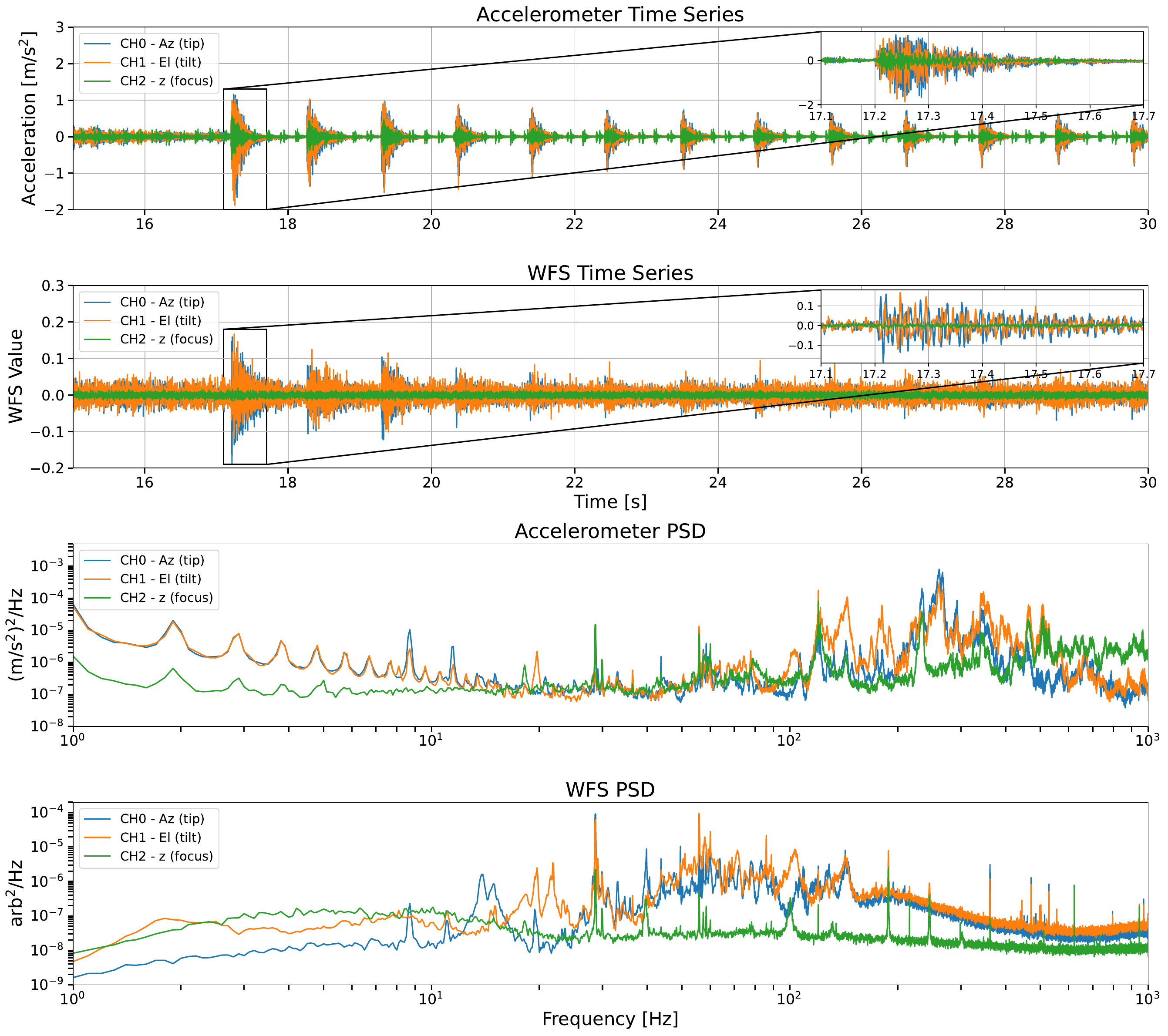}
\caption{
Time series and PSDs showing a strong 1 Hz feature introduced by the secondary mirror actuation system controlling secondary position.
}
\label{fig:tele_ts_psds}
\end{figure}

Figure~\ref{fig:tele_ts_psds} shows time series from the accelerometer and the WFS. The periodic 1 Hz signal is caused by the vane end actuation system which controls the position of the secondary, with five degrees of freedom, and the tension in the vane end. In the telescopes autofocus mode, this system was being enabled every second whether a change in position was required or not. The control software has since been updated so that this system is activated only when a position correction is required. In addition, within the MagAO-X operation procedure, we are able to coordinate with the telescope operator to run the telescope without autofocus on because slow focus drifts can be compensated using the MagAO-X woofer DM \footnote{\url{https://magao-x.org/docs/handbook/}}.

We observe low-frequency spikes in the accelerometer PSD, as shown in Figure~\ref{fig:tele_ts_psds}, and analyzed this in simulation which suggests that these features can be explained by circuit-induced offsets that mimic a low-amplitude square wave. This is caused by the circuit having more dynamic range below the midpoint of the ADC digital numbers. Therefore, when the large amplitude 1 Hz spikes occur in the timeseries, the positive accelerations are clamped by the circuit sooner than the negative accelerations.

\subsection{Accelerometer and WFS Correlation}

\begin{figure}[!ht]
\centering
\includegraphics[width=\columnwidth]{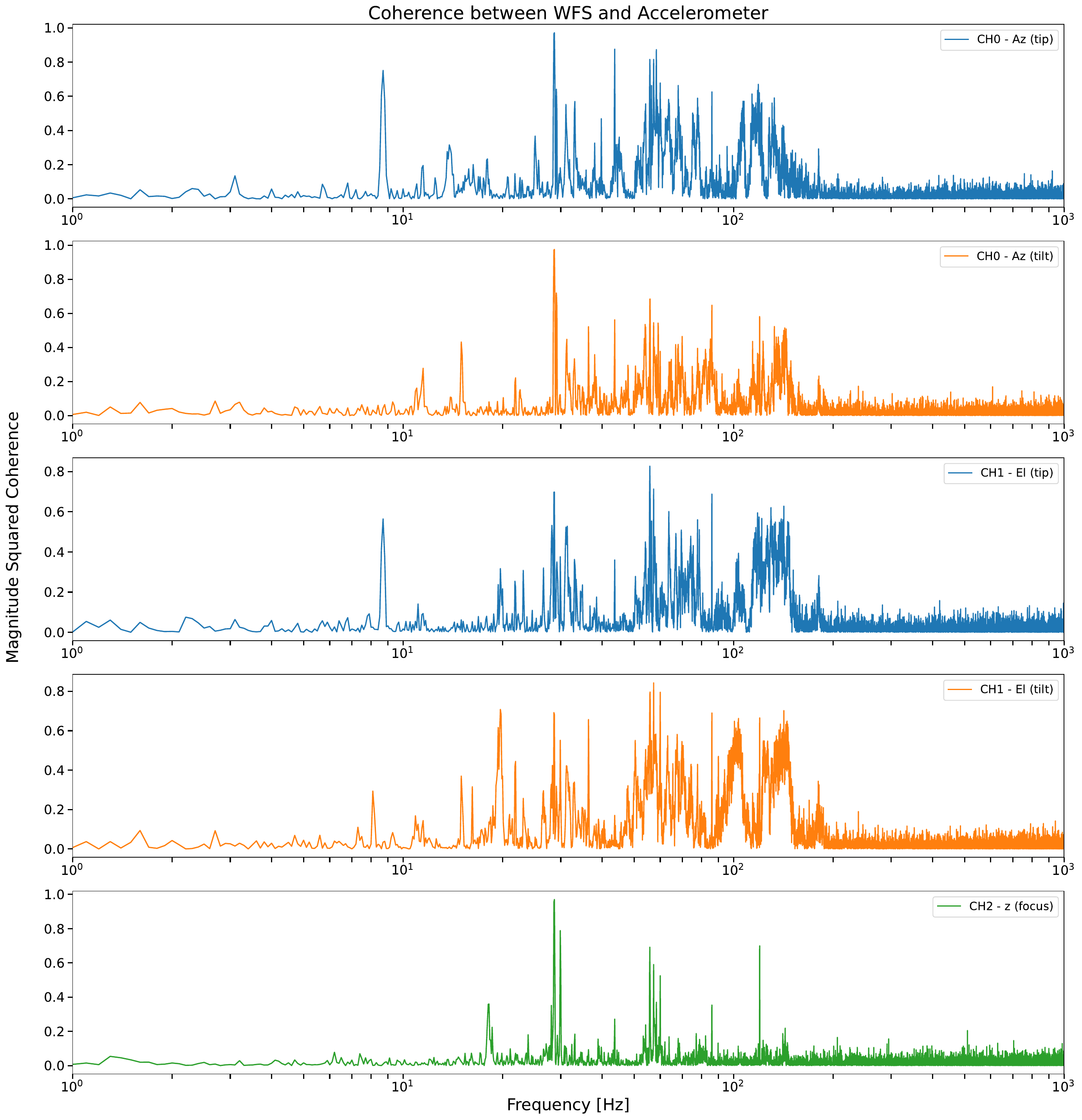}
\caption{
Magnitude-squared coherence between accelerometer and WFS signals, where values range from 0 (no correlation) to 1 (perfect correlation at a given frequency).
}
\label{fig:coherence}
\end{figure}

To estimate the fraction of the WFS signal that is correlated with structural vibrations, we compute the magnitude-squared coherence between the accelerometer and WFS telemetry streams. Figure~\ref{fig:coherence} shows the magnitude-squared coherence between the azimuth and elevation accelerations with the tip and tilt modes on the WFS, as well as the z direction accelerometer with focus. We can use this coherence to determine the percentage the accelerometer signals account for the low-order WFS modes. To quantify the fraction of WFS power associated with structural
vibrations, we define the coherence-weighted explained power as,
\begin{equation}
P_{\mathrm{exp}} = \int_{0}^{\infty} C_{xy}(f)\, P_{yy}(f)\, df,
\end{equation}
where $C_{xy}(f)$ is the magnitude-squared coherence between the accelerometer signal $x(t)$ and the WFS signal $y(t)$ at frequency $f$, and $P_{yy}(f)$ is the PSD of the WFS signal. The total WFS power is given by,
\begin{equation}
P_{\mathrm{tot}} = \int_{0}^{\infty} P_{yy}(f)\, df.
\end{equation}
The fraction of WFS power explained by the accelerometer measurements is then defined as,
\begin{equation}
\eta = \left( \frac{ \int_{0}^{\infty} C_{xy}(f)\, P_{yy}(f)\, df }{ \int_{0}^{\infty} P_{yy}(f)\, df } \right) \times 100.
\end{equation}
This provides a frequency weighted estimate of the coupling between accelerometer measurements and WFE. Figure~\ref{fig:coherence} reveals the frequencies that have the highest correlation between the accelerometers and WFS, and Table~\ref{tab:explained_power} summarizes the explained power between each channel.

\begin{table}[ht]
\centering
\caption{Fraction of WFS power explained by accelerometer measurements.}
\label{tab:explained_power}
\begin{tabular}{l c}
\hline
Channel & Explained Power [$\%$] \\
\hline
CH0 (Azimuth, Tip)   & 26.29 \\
CH0 (Azimuth, Tilt)  & 19.79 \\
CH1 (Elevation, Tip) & 22.99 \\
CH1 (Elevation, Tilt)& 31.05 \\
CH2 (z, Focus)          & 5.01 \\
\hline
\end{tabular}
\end{table}

To estimate the net accelerometer-correlated contribution in each physical axis, the tip and tilt contributions are combined in quadrature, assuming that the residual tip and tilt components are approximately statistically independent,
\begin{equation}
\eta_{\mathrm{Az}} = \sqrt{\eta_{\mathrm{tip}}^2 + \eta_{\mathrm{tilt}}^2} =
32.91\%,
\end{equation}
\begin{equation}
\eta_{\mathrm{El}} = \sqrt{\eta_{\mathrm{tip}}^2 + \eta_{\mathrm{tilt}}^2} =
38.64\%.
\end{equation}
These results suggest that roughly one third of the residual tip/tilt variance is correlated with structural vibrations. This highlights the potential benefit of incorporating accelerometer measurements into predictive AO control algorithms.

\section{Discussion}
\label{sec:discussion}

\subsection{GMT Implications}
\label{sec:gmt_implication}

This work has important implications for future ELTs and their AO systems, particularly the GMT and GMagAO-X. ELTs are expected to experience significantly stronger structural vibrations due to increased mechanical complexity, larger moving masses, and greater wind forces on telescope structures\cite{Jaufmann24,Jaufmann25}. In addition, larger secondary mirrors, support structures, cooling systems, pumps, and rotating mechanisms will introduce additional vibration modes that will propagate directly into the optical path.

For HCI instruments such as GMagAO-X\cite{Males24, Close26, Males26}, these disturbances will directly limit coronagraphic performance if not identified and eliminated or actively controlled. We have shown mechanical vibrations produce low-order aberrations that reduce coronagraphic suppression and increase stellar leakage. Since ELT science goals rely on detecting faint companions and circumstellar structure at extremely small angular separations, vibration mitigation should be considered a fundamental component of both GMagAO-X and GMT system design, rather than a secondary optimization.

\subsection{Limitations}
\label{sec:limitations}

Several limitations remain in our current implementation. The accelerometers only measure local structural motion, meaning they cannot sense all vibrations contributing to WFE. Vibrations that do not couple strongly to the chosen sensor mounting location will remain unobserved. Because the acquisition system is designed to be modular and scalable, future work will investigate optimal sensor placement and incorporate additional accelerometers throughout the telescope structure to improve coverage of vibration sources that couple into the optical path.

While this work demonstrates synchronized vibration telemetry and identifies structural vibration sources, integrating accelerometer measurements into on-sky AO real-time control systems remains future work. Future studies will investigate predictive control implementations using the synchronized telemetry presented here.

\section{Conclusion}
\label{sec:conclusion}

This work demonstrates that low cost accelerometer systems can be integrated into an AO telemetry pipeline and used to directly measure structural vibrations correlated with WFE. The results show that a significant fraction of the low-order WFS signal is driven by mechanical disturbances rather than atmospheric turbulence alone. We were able to identify dominant vibration sources, quantify their coupling to the WFS, and demonstrate that synchronized accelerometer telemetry provides a promising approach for future accelerometer-assisted predictive control in current and future AO systems.

An important outcome of this work is the successful implementation of a low-latency acquisition and synchronization system. The direct Ethernet connection between the Raspberry Pi acquisition system and the RTC provides sufficiently low latency for predictive control algorithms while remaining inexpensive and modular. CPU isolation reduced timing jitter to only a few microseconds RMS, well below the WFS integration time. This level of synchronization is critical because predictive control performance depends strongly on temporal alignment between accelerometer measurements, WFS telemetry, and DM commands.

The accelerometer measurements also proved effective for identifying vibration sources throughout the telescope system. Multiple narrow-band vibration peaks observed in the WFS residuals were directly correlated with telescope subsystems, including the primary mirror glycol pump, secondary mirror actuation system, and the autofocus system. In several cases, operational changes implemented by the MagAO-X team and Las Campanas Observatory staff reduced vibration amplitudes without requiring hardware modifications or predictive control algorithms. For example, disabling unnecessary actuation system updates eliminates the strong 1 Hz feature observed in both the accelerometer and WFS telemetry. Also, operating MagAO-X with the telescope autofocus system disabled reduced the high-frequency disturbances observed between 100 and 500 Hz.

Overall, this work demonstrates that low-cost synchronized accelerometer telemetry provides an effective framework for identifying structural vibration sources, quantifying their coupling to residual WFE, and guiding vibration mitigation strategies for current and future HCI systems to maximize scientific return.

\appendix    

\acknowledgments 

This material is based on work supported by the National Science Foundation (NSF) Graduate Research Fellowship Program under grant No. 2025372994. Any opinions, findings, and conclusions or recommendations expressed in this material are those of the author(s) and do not necessarily reflect the views of the NSF. We are very grateful for support from the NSF MRI Award No. 1625441. The Phase II upgrade program is made possible by the generous support of the Heising-Simons Foundation. MagAO-X uses the CACAO software package, which is supported by NSF Award No. 2410616. This work made use of the astropy, SciPy, and numpy software packages. We also would like to acknowledge the support of all the Las Campanas Observatory's staff and engineers that made the hardware implementation possible. 

\bibliography{report} 
\bibliographystyle{spiebib} 

\end{document}